\documentclass[aps,prl,twocolumn,amsmath,preprintnumbers,nofootinbib,a4paper]{revtex4-1}

\usepackage{amsthm}
\usepackage{amssymb}
\usepackage{amsfonts}
\usepackage{bbm}
\usepackage{color}
\usepackage{dcolumn}
\usepackage{bm}
\usepackage{pxfonts}
\usepackage{slashed}
\usepackage{graphicx}
\usepackage{multirow}

\usepackage[bookmarks]{hyperref}
  
\newcommand{\beq}{\begin{eqnarray}}
\newcommand{\eeq}{\end{eqnarray}}

\newcommand{\bmp}{\noindent\begin{minipage}{16cm}}
\newcommand{\emp}{\end{minipage}\vskip 7mm} 

\definecolor{bluc}{cmyk}{1,1,0,0.1}
\definecolor{rossoCP3}{cmyk}{0,.88,.77,.40}
\definecolor{rosso}{cmyk}{0,1,1,0.4}
\definecolor{rossos}{cmyk}{0,1,1,0.55}
\definecolor{rossoc}{cmyk}{0,1,1,0.2}
\definecolor{verdes}{cmyk}{0.92,0,0.59,0.4}

\hypersetup{colorlinks, bookmarksopen, bookmarksnumbered,
citecolor=verdes, linkcolor=bluc, pdfstartview=FitH, urlcolor=rossos}

\def\lsim{\mathrel{\rlap{\lower4pt\hbox{\hskip1pt$\sim$}}
    \raise1pt\hbox{$<$}}}                
\def\gsim{\mathrel{\rlap{\lower4pt\hbox{\hskip1pt$\sim$}}
    \raise1pt\hbox{$>$}}}                

\newcommand{\ba}{\begin{eqnarray}}
\newcommand{\ea}{\end{eqnarray}}
\newcommand{\be}{\begin{equation}}
\newcommand{\ee}{\end{equation}}
\newcommand{\bd}{\begin{displaymath}}
\newcommand{\ed}{\end{displaymath}}
\newcommand{\een}{\nonumber\end{equation}}
\newcommand{\bea}{\begin{eqnarray}}
\newcommand{\eean}{\nonumber\end{eqnarray}}
\newcommand{\eea}{\end{eqnarray}}

\def\eq#1{Eq.~(\ref{#1})}

\def\fig#1{Fig.~\ref{#1}}

\def\tab#1{Table~\ref{#1}}

\def\cite#1{\citep{#1}}

\begin{document}

\title{\Large  \color{rossoCP3} Anomalous Dimensions of Conformal Baryons}
\author{Claudio Pica$^{\color{rossoCP3}{\varheartsuit}}$}\email{pica@cp3-origins.net}
\author{Francesco Sannino$^{\color{rossoCP3}{\varheartsuit}}$}\email{sannino@cp3-origins.net}

\affiliation{
{$^{\color{rossoCP3}{\varheartsuit}}${ \color{rossoCP3}  \rm  CP}$^{\color{rossoCP3} \bf 3}${\color{rossoCP3}\rm-Origins} \& the {\color{rossoCP3} \rm Danish IAS}, University of Southern Denmark, Campusvej 55, DK-5230 Odense M, Denmark \\}
 }
\begin{abstract}
We determine the anomalous dimensions of baryon operators for the three color theory as function of the number of massless flavours within the conformal window to the maximum known order in perturbation theory. We show that the anomalous dimension of the baryon is controllably small, within the $\delta$ expansion, for a wide range of number of flavours. We also find that this is always smaller than the anomalous dimension of the fermion mass operator. These findings challenge the partial compositeness paradigm.
 \end{abstract} 
 \preprint{CP3-Origins-2016-019 DNRF90}
 \preprint{DIAS-2016-19}
 \maketitle


Determining the phase structure of  gauge theories of fundamental interactions is crucial when selecting relevant extensions of the standard model  \cite{Sannino:2009za}. Of particular significance are the critical exponents of composite conformal operators such as the fermion mass and the baryon anomalous dimensions in the conformal window. 
Large anomalous dimensions of these operators are often invoked when constructing composite extensions of the standard model, such as models of walking dynamics \cite{Holdom:1984sk} and partial compositeness \cite{Kaplan:1991dc}. 

To gain a quantitative analytic understanding of these important quantities perturbation theory has been proven useful for the anomalous dimension of the fermion bilinear \cite{Pica:2010mt,Ryttov:2010iz,Pica:2010xq}.
 Here we determine the conformal baryon anomalous dimension for the SU($3$) gauge theory when varying the number of massless flavours within the conformal window to the maximum known order in perturbation theory.
 These operators play an important role in models of partial compositeness.

\section{Baryon anomalous dimension}
The perturbative expressions of the beta function and the fermion mass anomalous dimension for a generic gauge theory with only fermionic matter in the $\overline{\rm MS}$ scheme to four loops were derived in \cite{vanRitbergen:1997va,Vermaseren:1997fq} and assume the general form: 
\begin{eqnarray}
\label{beta}
\frac{d a }{d \ln \mu^2}  & = &
\beta(a) = -\beta_0 a^2 - \beta_1 a^3
-\beta_2 a^4
-\beta_3 a^5 + O(a^6)\ , \hspace{.5cm}
\end{eqnarray}
\begin{eqnarray}
 \label{gamma}
-\frac{d\ln m }{ d \ln \mu^2}  & = & 
\frac{\gamma_m (a)}{2}  =  \gamma_0 a +  \gamma_1 a^2
+ \gamma_2 a^3
+ \gamma_3 a^4 + O(a^5) \ ,\hspace{.5cm}
\end{eqnarray}
where $m=m(\mu)$ is the renormalized (running)  fermion mass, 
$\mu$ is  the renormalization scale in the $\overline{\rm MS}$ scheme and
$a=\alpha/4\pi=g^2/16\pi^2$ where $g=g(\mu^2)$ is the renormalized 
coupling constant  of the theory. The explicit four loop values of the coefficients for generic fermion representations were generalised from the original references in \cite{Mojaza:2010cm} while the explicit formulae for the coefficients are shown in the appendix of \cite{Pica:2010xq}.  
 
We consider an SU($3$) gauge theory with $N_f$ fundamental massless Dirac flavours.   Because of the triality condition, as for QCD, the baryon operator is constructed out of three quarks. In general for the renormalization of composite operators one needs to consider operator mixing.  This results in a matrix of 
renormalization constants and related matrix of anomalous dimensions.   The renormalized baryon operators are given in terms of the bare ones ${\cal B}^{\,b}_j$ according to the standard relation  \begin{equation}
{\cal B}_{i} = Z_{ij} {\cal B}^{\,b}_j \ ,
\end{equation}
where the indices $i,j$ range over all operators that mix.
The matrix of anomalous dimensions are given by
\begin{equation}
\gamma_{ij}^{\cal B} (a) = \mu \frac{d ~}{d \mu} \ln Z_{ij}\, .
\label{matdef}
\end{equation}
In \cite{Pivovarov:1991nk,Gracey:2012gx} the anomalous dimensions of the proton-like baryon was derived considering the mixing of two three-quarks operators.  Using the same notation as in \cite{Gracey:2012gx}, the resulting three-loop eigen-anomalous dimensions are:
\begin{widetext}
 \begin{eqnarray}
\gamma_+^{\cal B}(a) &=& 2 a + \left[ 2 n_f + 21 \right] \frac{a^2}{9}
-\left[ 260 n_f^2 + [ 4320 \zeta(3) - 4740 ] n_f + 2592 \zeta(3) + 22563 
\right] \frac{a^3}{162} +~ O(a^4)\, , \nonumber \\
\gamma_-^{\cal B}(a) &=&  2 a + \left[ 2 n_f + 81 \right] \frac{a^2}{9}
- \left[ 260 n_f^2 + [ 4320 \zeta(3) - 4572 ] n_f + 24399 \right] 
\frac{a^3}{162}  +~ O(a^4) \, .
\label{BaryonAD}
\end{eqnarray}
\end{widetext}

In the conformal window the anomalous dimensions at the infrared (IR) fixed point are physical quantities, which do not depend on the scheme. At fixed loop order, the scheme independence has been studied in \cite{Ryttov:2014nda}.

\begin{figure}[t!]
\includegraphics[width=\columnwidth]{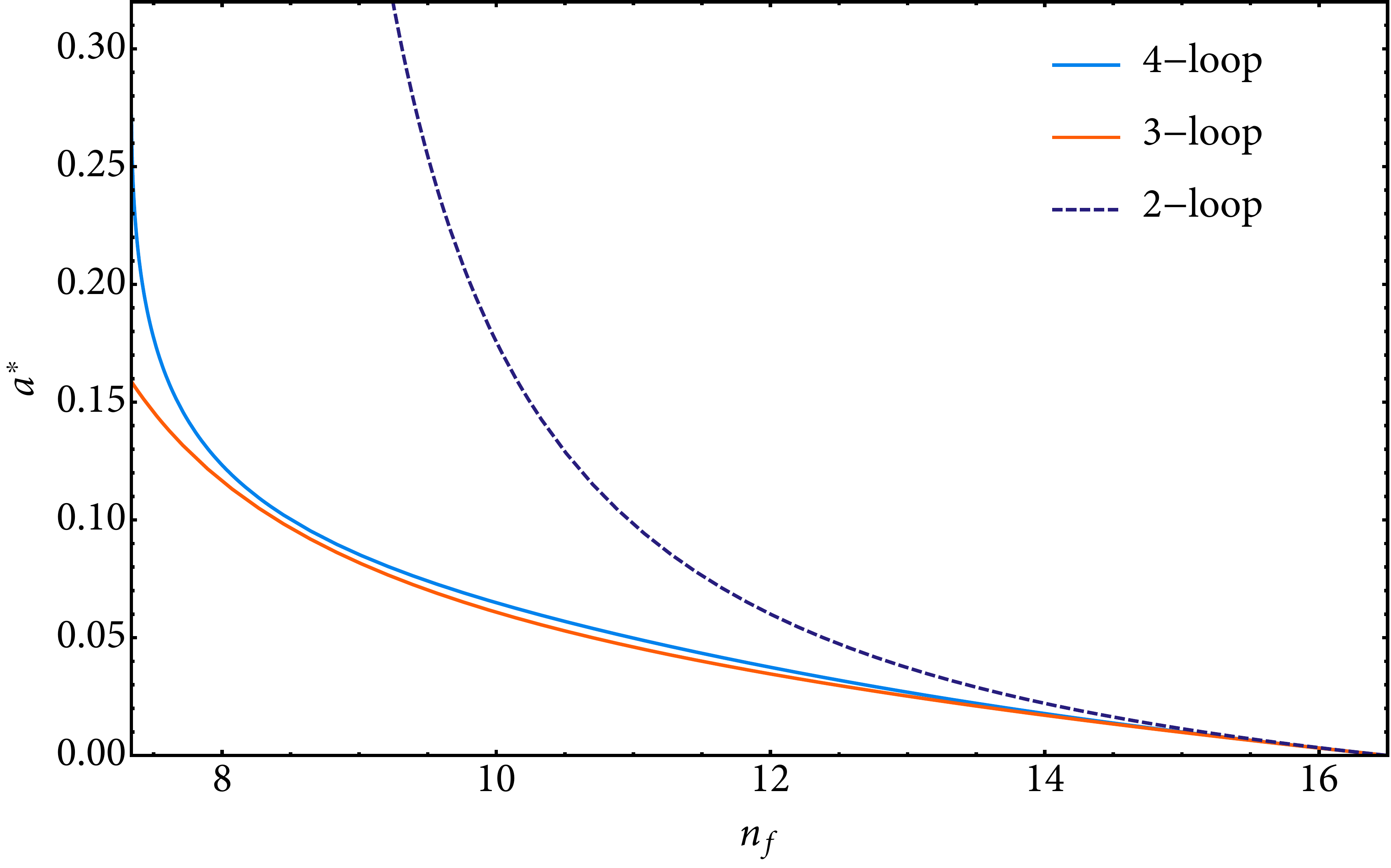}
\caption{Value of the fixed point coupling $a^*$ at 4-, 3- and 2-loop order as a function of the number of flavors $n_f$. The 4-loop fixed point coupling reaches a finite value at the lower end of the 4-loop conformal window at $n_f\approx 7.3$, while both the 3- and 2-loop fixed point couplings diverge there. \label{gcrit}}
\end{figure}

Using the known expression for the $\beta$-function \eq{beta}, one can determine the value of the IR fixed point coupling $a^*$ at 2-, 3- and 4-loops, which we show in \fig{gcrit} \cite{Ryttov:2010iz,Pica:2010xq}. From the figure it is clear that the 2-loop result is only accurate very close to the upper limit of the conformal window, i.e. $n_f\ge 14$, while the agreement between the 3- and 4-loop result suggests that the fixed point coupling so determined is reliable for a much wider range of number of flavors, i.e. $n_f\gsim 8$. 

\begin{figure}[t!]
\includegraphics[width=\columnwidth]{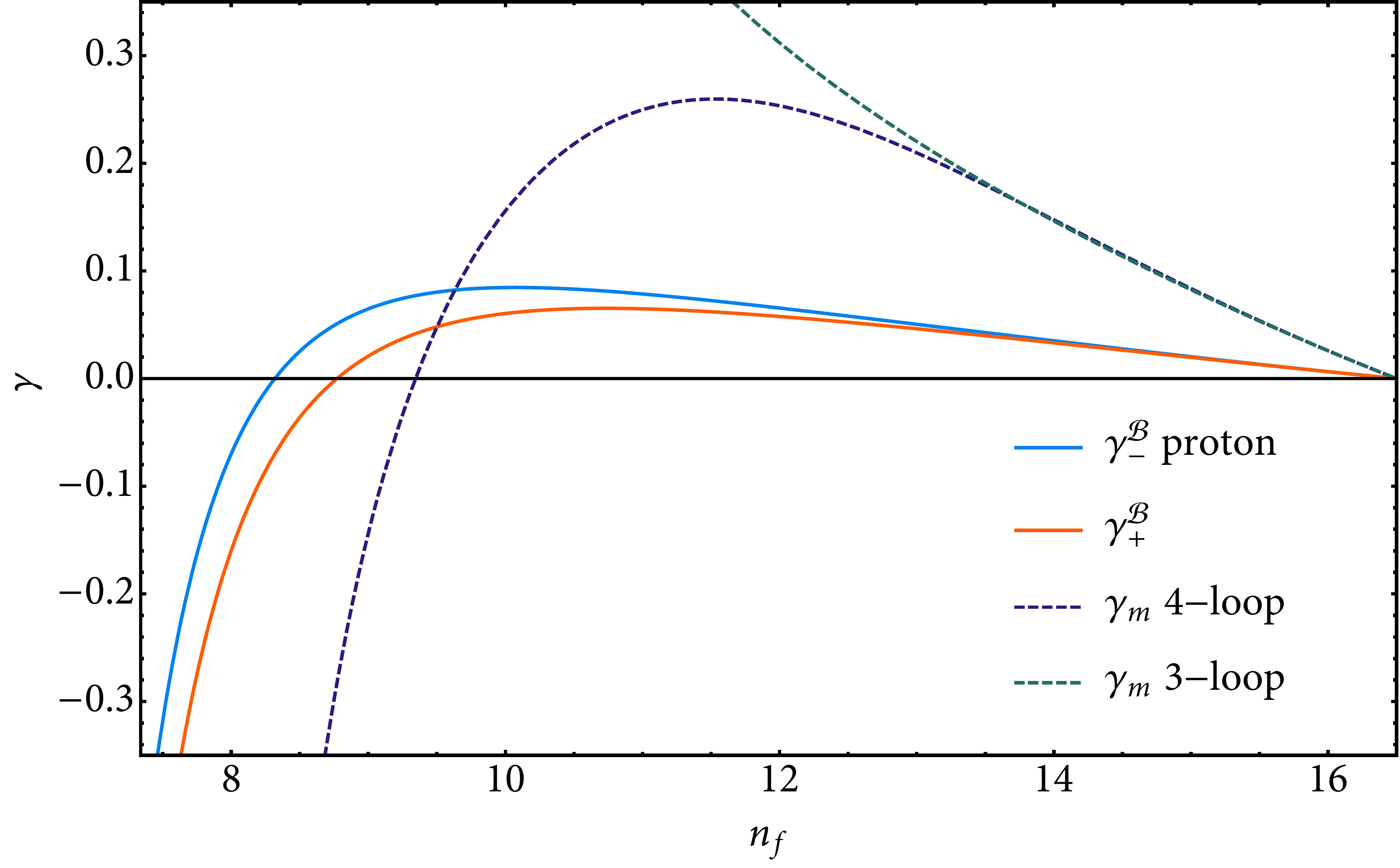}
\caption{Eigen-anomalous dimensions $\gamma_\pm^{\cal B}$ for 3-quark composite operators compared to the anomalous dimension of the mass $\gamma_m$ at 3- and 4-loops as a fucntion of $n_f$. The anomalous dimesion $\gamma_-$ corresponds to the proton in QCD. \label{fig:gamma}}
\end{figure}

Using the value of the fixed point coupling $a^*$, we can now determine the anomalous dimensions of baryons $\gamma_\pm^{\cal B}$.  In \fig{fig:gamma} we compare the 3-loop result for $\gamma_\pm^{\cal B}$ with the 3- and 4-loop results for the mass anomalous dimension $\gamma_m$ at the IR fixed point. 
The range of validity of perturbation theory for $\gamma_m$ can be estimated by comparing the 3- and 4-loops result, which suggests $n_f\ge 12$ \cite{Ryttov:2010iz,Pica:2010xq}. In fact, the anomalous dimension of the fermion mass term  $\gamma_m$ have also been investigated on the lattice by several groups and for the purpose of this work we refer to the comprehensive review of these results presented in \cite{Giedt:2015alr}. Although no universal consensus exists yet on whether or not the SU(3) theory with $n_f = 12$ has IR fixed point~\cite{lattice}, the lattice results measuring the anomalous mass dimension are compatible with the four loop results. This is in agreement with the expectation that perturbation theory is valid till $n_f = 12$.

\begin{figure}[t!]
\includegraphics[width=\columnwidth]{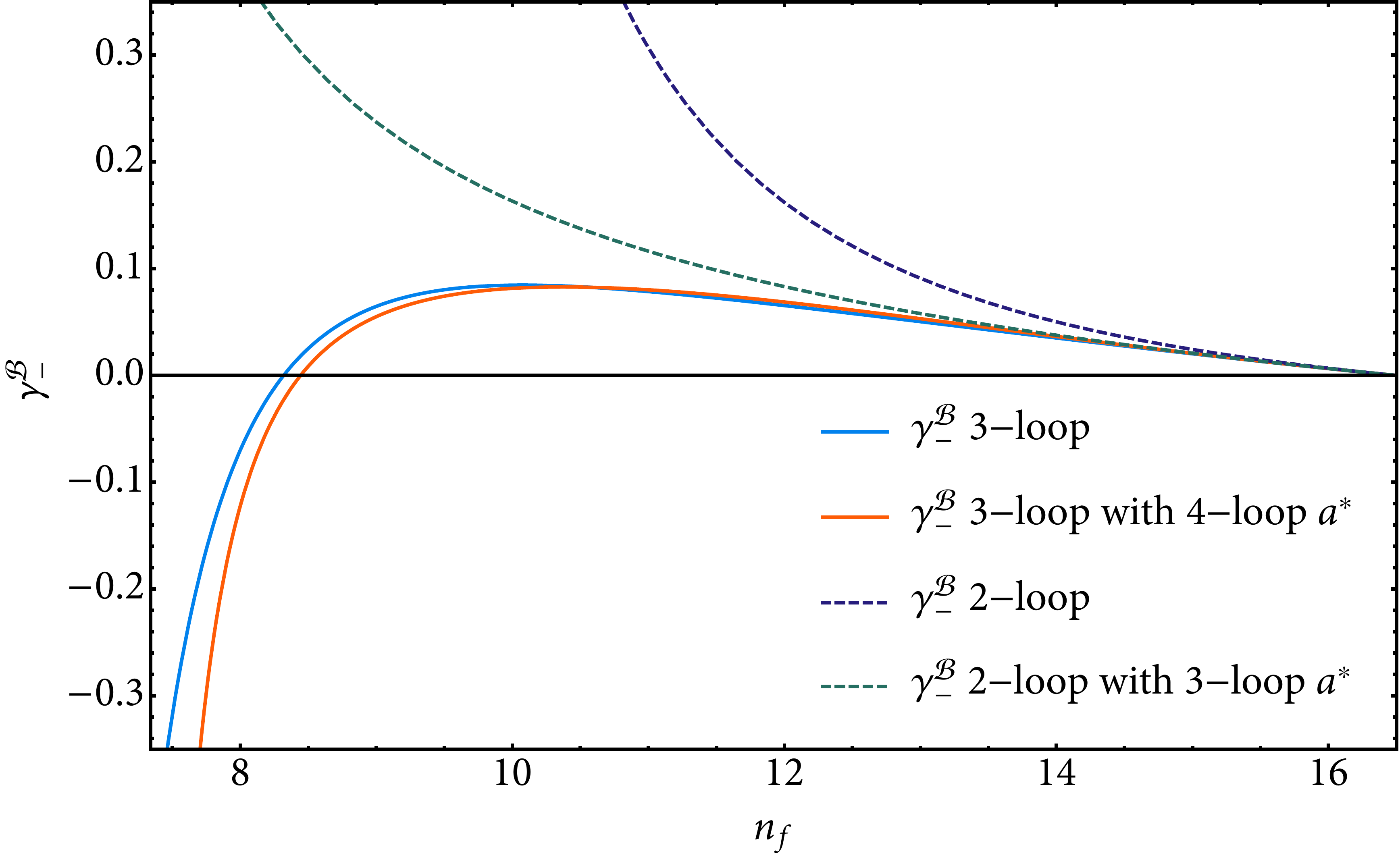}
\caption{To estimate the convergence of the perturbative series for $\gamma_-^{\cal B}$, we evaluate its 2- and 3-loop expression using the 2-, 3- and 4-loop values of the fixed point value $a^*$.  \label{gproton}}
\end{figure}

In this range, the baryon anomalous dimension is very small $\le 0.07$, about a factor four smaller than the mass anomalous dimension. It is also apparent that the two eigen-anomalous dimensions of the conformal baryon are, perhaps not surprisingly, very close to each other.
To investigate the stability of the perturbative result  for $\gamma_-^{\cal B}$, we compare in \fig{gproton} various perturbative estimates. We use the 3- and 2- loop expressions for $\gamma_-^{\cal B}$ and for each we insert the value of the IR fixed point coupling at the same order in perturbation theory or one order higher. For $n_f\ge 12$, we observe a good agreement among three of these four estimates,  corresponding to the  3-loop $\gamma_-^{\cal B}$ computed at the 3- and 4-loop value of $a^*$ and the 2-loop $\gamma_-^{\cal B}$ computed at the 3-loop value of $a^*$. 
The 2-loop estimate of the baryon anomalous dimension shows significant deviations respect to the other three estimates.This is consistent with the expectation that the 2-loop computation is reliable only very close to the loss of asymptotic freedom.
We report in \tab{table:gamma} the perturbative values of the anomalous dimensions determined in this work as a function of the number of flavors.

\begin{table}[t!]
  \begin{tabular}{r|ccccccc}
 
 $n_f$ & 10 & 11 & 12 & 13 & 14 & 15 & 16 \\
    \hline
    $\gamma_m$ & 0.1559 & 0.2497 & 0.2533 & 0.2098 & 0.1474 & 0.0836 & 0.0259 \\
 $\gamma_-^{\cal B}$&0.0816 & 0.0802 & 0.0688 & 0.0531 & 0.0365 & 0.0207 & 0.0064 \\
 $\gamma_+^{\cal B}$&0.0542 & 0.0641 & 0.0597 & 0.0484 & 0.0344 & 0.0200 & 0.0064 
  \end{tabular}
  \caption{Anomalous dimension of the mass $\gamma_m$ and of 3-quark operators $\gamma_\pm^{\cal B}$ as a function of the number of fundamental Dirac fermions $n_f$ to three loops order.\label{table:gamma}}
\end{table}

The physical dimension of the conformal baryon is given by:
\begin{equation}
  {\cal D}[{\cal B}_\pm]=\frac92-\gamma_\pm^{\cal B}\, ,
\end{equation}
and therefore it will remain very close to the engineering dimension $9/2$ for $n_f\ge 12$.

\section{Approaching the lower boundary of the conformal window}

\begin{figure}[t!]
\includegraphics[width=\columnwidth]{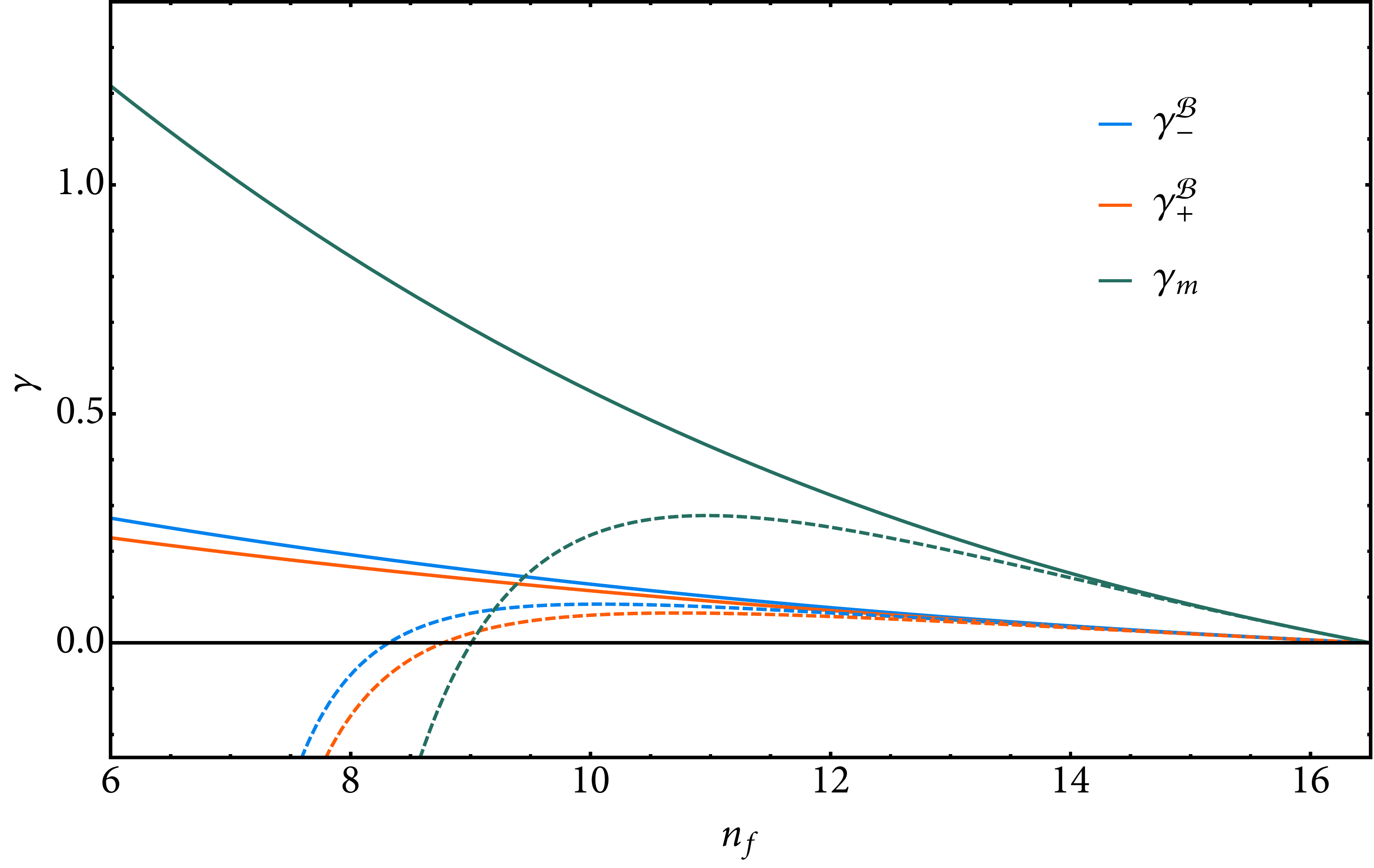}
\caption{Anomalous dimensions of the quark mass and of the 3-quark operators $\gamma_\pm^{\cal B}$ as a function of the number of fundamental Dirac fermions $n_f$ in the physical $\delta = n_f^{\rm{AF}}  - n_f$ expansion (solid lines). We compare it with the perturbative results (dashed lines) at three loops.\label{fig-dexp}}
\end{figure}

It is also possible to expand the anomalous dimensions at the fixed point in terms of the physical parameter $\delta = n_f^{\rm{AF}}  - n_f$ where $n_f^{\rm AF} = 16.5 $ is the value for which asymptotic freedom is lost \cite{Banks:1981nn,Pica:2010mt,Ryttov:2016hdp}. 
It is worth summarising the properties of the $\delta$ expansion \cite{Pica:2010mt,Ryttov:2016hdp}: this is the expansion of a physical quantity, here the anomalous dimensions at the IR fixed point, in terms of the physical parameter $\delta$ around the point where asymptotic freedom is lost; the n{\it th} coefficient of the series can be computed from the perturbative expansion at n{\it th}+1 loop order, but it is exact to all higher orders.
Furthermore, another virtue of this expansion is that it offers a more immediate evidence of scheme independence \cite{Pica:2010mt,Ryttov:2016hdp} and, as tested in \cite{Ryttov:2016hdp},  it converges rapidly   in the entire conformal window, to the exact results in supersymmetry.  We observe a similar convergence pattern for the coefficients of the series in $\delta$ also for the conformal baryon anomalous dimensions. The approximated numerical expressions, with rounded coefficients, are: 
\begin{eqnarray}
\gamma_-^{\cal B} & = & 0.012461\,\delta + 0.000845 \,\delta^2 + 0.000042 \, \delta^3 + {\cal O}(\delta^4) \ ,\nonumber \\
\gamma_+^{\cal B} & = & 0.012461\,\delta + 0.000586 \,\delta^2 + 0.000029 \, \delta^3 + {\cal O}(\delta^4) \ . \nonumber \\ 
\end{eqnarray}
 We plot in Fig.~\ref{fig-dexp} the conformal baryon anomalous dimensions in the $\delta$-expansion and compare them with the results presented above. 

From the analysis above clearly the anomalous dimensions are largest for lower number of flavours. It is therefore relevant to know the location of the lower end of the conformal window. This is being investigated by several lattice groups  \cite{Deuzeman:2008sc,Appelquist:2009ty,Fodor:2011tu,Aoki:2013xza,Appelquist:2014zsa,Aoki:2014oha,Cheng:2013eu,Hasenfratz:2014rna,Fodor:2015baa,daSilva:2015vna}. However, at present there is no general consensus on the location of this lower boundary: even if most studies indicate a number of flavours around eight, there are claims that it could be as large as twelve flavours~\cite{Fodor:2011tu}.

If the conformal window extends down to six flavours we have argued that the $\delta$ expansion converges down to this value with the baryon anomalous dimensions never exceeding the value of about 0.3.  

Another possible estimate of the lower boundary of the conformal window can be made by observing when the anomalous dimension of the mass operator approaches unity. Within the $\delta$ expansion, at order $\delta^3$, this occurs around seven flavours for which $\gamma_m \simeq 1.02$ \cite{Ryttov:2016hdp}. We report in Table~\ref{table:dgamma} the values of the anomalous dimensions considered in this work computed in the $\delta$-expansion.

\section{Phenomenological Impact}
Models of partial compositeness typically require the presence of baryonic operators with large anomalous dimensions such that 
\begin{equation}
  \frac32\le {\cal D}[{\cal B}]\le\frac52\, .
\end{equation}
This implies, for models in which the baryonic operators are composites of  three fermions \cite{Ferretti:2013kya,Barnard:2013zea,Ferretti:2014qta,Vecchi:2015fma,Kaplan:1991dc}, that the anomalous dimension should be
\begin{equation}
  2\le \gamma^{\cal B}\le 3\, .
\end{equation}

Our results indicate that such large anomalous dimensions can hardly occur in the minimal template of an SU($3$) model with $n_f$ fundamental fermions.


Furthermore lattice computations for the anomalous dimension of the mass \cite{Giedt:2015alr}, find no solid evidence of large anomalous mass dimensions suggesting that these might not be generated at the lower boundary of the conformal window even for fermions in higher representations.

Partial compositeness also requires four-fermion operators to be less relevant with respect to baryonic operators. 
By estimating the anomalous dimension of  four-fermion operators ${\cal D}[(\bar\psi\psi)^2]\approx 6-2\gamma_m$  \cite{Holdom:1995fu}.  Therefore if $\gamma^{\cal B}\le\gamma_m$, as in the present case, four-fermion operators will always be more relevant than baryonic ones. This estimate became a precise in the weakly coupled limit.

\begin{table}[t!]
  \begin{tabular}{l|ccc} 
$n_f$ & $\gamma _m$ & $\gamma _-^{\mathcal{B}}$ & $\gamma _+^{\mathcal{B}}$ \
\\\hline
 6 & 1.2160 & 0.2725 & 0.2295 \\
 7 & 1.0190 & 0.2305 & 0.1965 \\
 8 & 0.8435 & 0.1927 & 0.1663 \\
 9 & 0.6874 & 0.1586 & 0.1388 \\
 10 & 0.5495 & 0.1282 & 0.1138 \\
 11 & 0.4284 & 0.1011 & 0.0912 \\
 12 & 0.3227 & 0.0770 & 0.0706 \\
 13 & 0.2311 & 0.0558 & 0.0521 \\
 14 & 0.1520 & 0.0371 & 0.0353 \\
 15 & 0.0841 & 0.0207 & 0.0201 \\
 16 & 0.0259 & 0.0064 & 0.0064 \\
   \end{tabular}
  \caption{Anomalous dimension of the mass $\gamma_m$ and of 3-quark operators $\gamma_\pm^{\cal B}$ as a function of the number of fundamental Dirac fermions $n_f$ in the $\delta$-expansion.\label{table:dgamma}}
\end{table}

\section{Conclusions}
We determined the anomalous dimension of baryonic operators for the SU($3$) gauge theory with $n_f$ fundamental fermions inside the conformal window at the maximum known order in perturbation theory.
Within the conformal window at $n_f\ge 12$ our result indicate that anomalous dimensions for baryons remain small  $\gamma^{\cal B}\le 0.07$, and substantially smaller, about a factor $\approx 4$, than the mass anomalous dimension.

We have argued that the physical $\delta$-expansion gives a rapidly converging expansion, which allows to obtain a reliable estimate for a number of flavors as low as $n_f=6$.
Also within the $\delta$-expansion, the anomalous dimensions of the baryons never exceed $0.3$ for $n_f$ as low as six.

Our results are obtained for the most minimal setup for partial compositeness. More involved constructions, requiring fermions in multiple representations, have been proposed. For example, we can consider composite baryons of an SU(4) gauge theory with five Majorana fermions in the 2-index antisymmetric representation and three Dirac fundamental fermions~\cite{Ferretti:2013kya}. In this model, the anomalous dimensions of composite baryon operators at one-loop have recently been computed~\cite{DeGrand:2015yna}. Comparing the one-loop coefficients of the baryon operators to the ones of the mass operators, we find that, in all cases $\gamma^{\cal B}\le\gamma_m$.

These results challenge minimal models of partial compositeness featuring baryon operators built out of three fermions.

\vspace{-.3cm}
\subsection{Acknowledgments}
\vspace{-.1cm}
We thank Gabriele Ferretti and Luca Vecchi for useful discussions.
We thank the Mainz Institute for Theoretical Physics (MITP)  for its kind hospitality and support during the meeting {\it Composite Dynamics: From Lattice to the LHC Run II}, 4-15  April  2016. 
This work was supported by the Danish National Research Foundation DNRF:90 grant and by a Lundbeck Foundation Fellowship grant.

\end{document}